\documentclass[onecolumn, 11pt, draftclsnofoot]{IEEEtran}

\usepackage{algorithm, algorithmicx, algpseudocode}
\usepackage{graphicx, amsmath}
\usepackage{enumerate}
\usepackage{psfrag}
\usepackage{color}
\usepackage{subfigure}
\usepackage{url}
\usepackage{amsfonts,amssymb}
\usepackage{array}
\usepackage{algorithm, algorithmicx, algpseudocode}
\usepackage{amssymb}
\usepackage{enumerate}

\newfont{\bb}{msbm10 scaled 1100}
\newcommand{\CC}{\mbox{\bb C}}

\newtheorem{proposition}{Proposition}

\DeclareMathOperator{\tr}{tr}

\DeclareMathOperator{\st}{subject\ to}

\begin{document}

\title{\LARGE Resource Allocation for Power Minimization in the Downlink of THP-based Spatial Multiplexing MIMO-OFDMA Systems}
\author{\vspace{-0.2cm}Marco~Moretti{,} \emph{Member, IEEE}, Luca~Sanguinetti, \emph{Member, IEEE}, Xiaodong~Wang, \emph{Fellow, IEEE} \thanks{\indent M. Moretti and L. Sanguinetti are with the University of Pisa,
Dipartimento di Ingegneria dell'Informazione, Italy (
\{marco.moretti,luca.sanguinetti\}@iet.unipi.it). 
X. Wang is with the Electrical Engineering Department of Columbia University, USA (wangx@ee.columbia.edu)}}
\maketitle
\vspace{-1.1cm}

\begin{abstract}
\vspace{-0.4cm}
In this work, we deal with resource allocation in the downlink of spatial multiplexing MIMO-OFDMA systems. In particular, we concentrate on the problem of jointly optimizing the transmit and receive processing matrices, the channel assignment and the power allocation with the objective of minimizing the total power consumption while satisfying different quality-of-service requirements. A layered architecture is used in which users are first partitioned in different groups on the basis of their channel quality and then channel assignment and transceiver design are sequentially addressed  starting from the group of users with most adverse channel conditions. The multi-user interference among users belonging to different groups is removed at the base station using a Tomlinson-Harashima pre-coder operating at user level. Numerical results are used to highlight the effectiveness of the proposed solution and to make comparisons with existing alternatives.
\end{abstract}
\vspace{-0.7cm}
\section{Introduction}
\vspace{-0.2cm}
Dynamic resource allocation  in multiple-input multiple-output (MIMO) systems based on orthogonal frequency-division multiple-access (OFDMA) technologies has gained considerable research interest \cite{Svensson2012}. In most cases, subcarriers are assigned to the active users in an exclusive manner without taking advantage of the multi-user diversity offered by the spatial domain. 
A possible solution {to exploit the spatial dimension} is to make use of space-division multiple-access (SDMA) schemes, {which allow the simultaneous transmission of different users over the same frequency band. The main impairment of SDMA is represented by multiple-access interference (MAI).  In downlink transmissions, MAI mitigation can only be accomplished at the BS using pre-filtering techniques.} The most common approach for interference mitigation is zero-forcing (ZF) linear beamforming, which relies on the idea of \emph{pre-inverting} the channel matrix at the transmitter.
Another approach is represented by the block-diagonalization ZF (BD-ZF) scheme originally proposed in \cite{Spencer2004}.
Particular attention has been also devoted to dirty paper coding (DPC) techniques \cite{Tran2010} even though their implementation is still much open. A possible solution in this direction is represented by Tomlinson-Harashima precoding (THP), which can be seen as a one dimensional DPC technique \cite{FisherBook} and has been widely used in the downlink of single-user and multi-user MIMO systems \cite{Stankovic2005}--\nocite {Zhou2006c,Sanguinetti2007}\cite{DAmico2008}. {In combination with pre-filtering, another way to deal with interference in SDMA-OFDMA systems is user partitioning, which basically consists in properly selecting the set of users transmitting on the same subcarriers. As illustrated in \cite{MaKlei2010}, a common approach is to first group together users whose channels have low spatial cross-correlation and then to assign the subcarriers to the various groups. In \cite{ZhaLeta2005}, the authors follow a completely different approach in which the users are first divided into groups such that the spatial cross-correlations among users in different groups is low as much as possible and then subcarriers are sequentially assigned within each group.} 

From the above discussion, it follows that the use of SDMA schemes in MIMO-OFDMA systems makes the problem of resource allocation more challenging as it requires the joint optimization of \emph{a}) channel assignment {and user partitioning}; \emph{b}) power allocation over all active links; \emph{c}) transmit and receive filters. To the best of our knowledge, there exists only a few works dealing with all the above problems {together}. 
In \cite{Ho2009}, {the authors employ BD-ZF and Lagrange dual decomposition} to derive a resource allocation scheme for minimizing the power consumption when individual user rate constraints are imposed. The main drawback of this approach is that an exhaustive search is required to find the best user allocation on each subchannel. A reduced complexity solution is illustrated in \cite{Hassan2009}, in which a two-step procedure is adopted to decouple BD-ZF beamforming from subcarrier and power allocation. Although simpler than \cite{Ho2009}, it still requires an exhaustive search over a subset of users. In \cite{Moretti2013}, the author exploits a layered architecture in which a user partitioning technique {(resembling that discussed in  \cite{ZhaLeta2005})} is first used in conjunction with BD-ZF to partially remove multiuser interference and then carrier assignment is performed jointly with transceiver design using a linear programming (LP) formulation of the allocation problem \cite{Kim2006.1}. 

In this work, we return to the layered architecture investigated in \cite{Moretti2013} and extend it in several directions. {First, we reformulate the power minimization problem assuming that the quality-of-service (QoS) constraint of each user is given as a sum of the mean-square-errors (MSEs) over all subcarriers rather than on the sum of the achievable rates. Second, transceiver design is carried out employing a non-linear THP precoder operating at \emph{user level} at the transmitter. Third, the choice of the user partitioning strategy is motivated by its combination with the THP precoding technique.} This allows us to completely remove the multiuser interference (rather than partially removing it) and to make use of a close-to-optimal partitioning strategy.  All this leads to a resource allocation scheme of affordable complexity, which is shown by means of numerical results to outperform the solution presented in \cite{Moretti2013}. 

\vspace{-0.3cm}
\section{Problem description}
We consider\footnote{We use ${\bf{A}} = {\rm{blkdiag}}\left\{
{{\bf{A}}_1,{\bf{A}}_2, \ldots ,{\bf{A}}_K} \right\}$ to
represent a block diagonal matrix whereas ${\bf{A}}^{ - 1}$ and
${\rm{tr}}\left\{ {\bf{A}} \right\}$ denote the inverse and trace of a
square matrix ${\bf{A}}$. We denote ${\bf{I}}_K$ the identity matrix of order $K$ while we use $ E\left\{ \cdot \right\}$ for
expectation, $\left\| \cdot \right\|$ for the Euclidean norm of
the enclosed vector and the superscript $ ^*$, $ ^T$ and $ ^H$ for
complex conjugation, transposition and Hermitian transposition. The
notation $\left[\cdot \right]_{k,\ell}$ indicates the ($k ,\ell$)th entry of the
enclosed matrix.} the downlink of an OFDMA network in which a total of $N$ subcarriers is used to communicate with $K$ MTs, each equipped with $N_R \geq 2$ antennas\footnote{{The results can be easily extended to a more versatile system in which a different number of services is required by each MT. In this case, $K$ would simply denote the total number of services.}}. The BS is endowed with $N_T > N_R$ transmit antennas. We denote by $\mathbf{s}_{n,k}$ the $N_T$-dimensional vector collecting the {data transmitted to} user $k$ on subcarrier $n$ and by $a_{n,k}\in \{0,1\}$ the binary allocation variable, which is equal to one if subchannel $n$ is assigned to user $k$ and zero otherwise. 
 The goal of this work is to minimize the total power consumption given by 
%
\begin{equation}\label{3.0}
P_T = \sum\limits_{n=1}^N \sum\limits_{k=1}^K{\rm{E}}\left\{\mathbf{s}_{n,k}^H\mathbf{s}_{n,k}\right\}
\end{equation}
while satisfying user QoS requirements given as a function of the sum of the MSEs over all their assigned subcarriers. To be more specific,
the expression for the $k$th user constraint is 
%
\begin{equation}\label{3.1}
\sum\limits_{n=1}^{N} a_{n,k}\sum\limits_{\ell=1}^L {\rm{MSE}}_{n,k} (\ell)\le \gamma_k
\end{equation}
where $L$ denotes the number of streams transmitted to the $k$th user over the $n$th subcarrier and ${\rm{MSE}}_{n,k} (\ell)$ denotes its corresponding MSE. The quantities $\gamma_{k} > 0$ are design parameters that specify different QoS requirements for each user.  We assume that a maximum number of $Q=\lfloor N_T/N_R\rfloor$ users can be simultaneously allocated over each subcarrier, so that  it is $\sum\nolimits_{k=1}^Ka_{n,k} \le Q$ for each channel $n$.
 To avoid the trivial solution where a user with no allocated subcarrier consumes no power and has a zero MSE, we require that at least $n_{k}$ {subcarriers} are assigned to each user so that it is $\sum\nolimits_{n=1}^Na_{n,k} \ge n_k $ $\forall k$.

\vspace{-0.3cm}
\section{Multi-user interference elimination and user partitioning}\label{THP}
{Unfortunately, solving the optimization problem described above requires an exhaustive search over all possible subcarrier allocations. Moreover, it needs also the joint optimization of the transmit and receive processing matrices for each allocation. All this makes its complexity extremely large for any practical scenario.
To address this issue, we follow the approach of \cite{ZhaLeta2005} and \cite{Moretti2013}, in which the population of $K$ users is partitioned into $Q$ different subsets $\{\mathcal S^{(1)}, \mathcal S^{(2)}, \ldots,\mathcal S^{(Q)}\}$. This allows us to break the original problem into a sequence of $Q$  lower-complexity optimization sub-problems, each assigning all radio resources to a subset of users. Users within the same subset are transmitted on orthogonal subcarriers  and do not interfere with each other. Channels allocation is performed sequentially starting from set $\mathcal S^{(1)}$.}
\par
{From the above discussion, it follows that, after the $Q$ allocation sub-problems are solved, there will be $Q$ users assigned to each subcarrier.  Without loss of generality, we focus on subcarrier $n$.} Let us denote  by $\mathcal K_n$ the set of users assigned to $n$ and by $\mu_n(i)$ the user in $\mathcal S^{(i)}$ associated to subcarrier $n$. To simplify the notation, in the following derivations the {indexes} $\mu_{n}(i)$ will be relabelled according to the map $\mu_{n}(i) \rightarrow i$.
The signal ${\bf{x}}_{n,k} \in \CC^{N_R \times 1}$ received at the $k$th MT over the $n$th subcarrier can be thus written as
\begin{equation}\label{eq:x_n,k.0}
{\bf{x}}_{n,k} = {\bf{H}}_{n,k} \sum\limits_{i=1}^Q\mathbf{s}_{n,i} + {\bf{w}}_{n,k} 
\end{equation}
where ${\bf{w}}_{n,k} \in \CC ^{N_{R} \times 1}$ is a Gaussian vector with zero-mean and covariance matrix $\sigma^2\mathbf{I}_{N_{R}}$ and ${\bf{H}}_{n,k} \in \CC ^{N_{R} \times N_T}$ is the channel matrix over the $n$th subcarrier. From \eqref{eq:x_n,k.0}, it follows that the interference term is given by two different contributions, namely, ${\bf{H}}_{n,k} \sum\nolimits_{i=1}^{k-1}\mathbf{s}_{n,i}$ and ${\bf{H}}_{n,k} \sum\nolimits_{i=k+1}^{Q}\mathbf{s}_{n,i}$. {The first term represents  the interference caused by the active users already allocated before the  $k$th assignment sub-problem has been solved (i.e., users belonging to sets $\mathcal S^{(i)}$ with indexes $i < k$), while the second term accounts for the users with indexes $i > k$ (i.e., users which have been allocated after user $k$).} In \cite{Moretti2013}, a BD-ZF scheme is employed to remove the first term while the second one is treated as Gaussian noise. In the sequel, a THP technique operating at user level is used to remove both terms.

\vspace{-0.4cm}
\subsection{Multi-user interference elimination}
\vspace{-0.1cm}

\begin{figure}[t]
\begin{center}
\includegraphics[width=.9\textwidth]{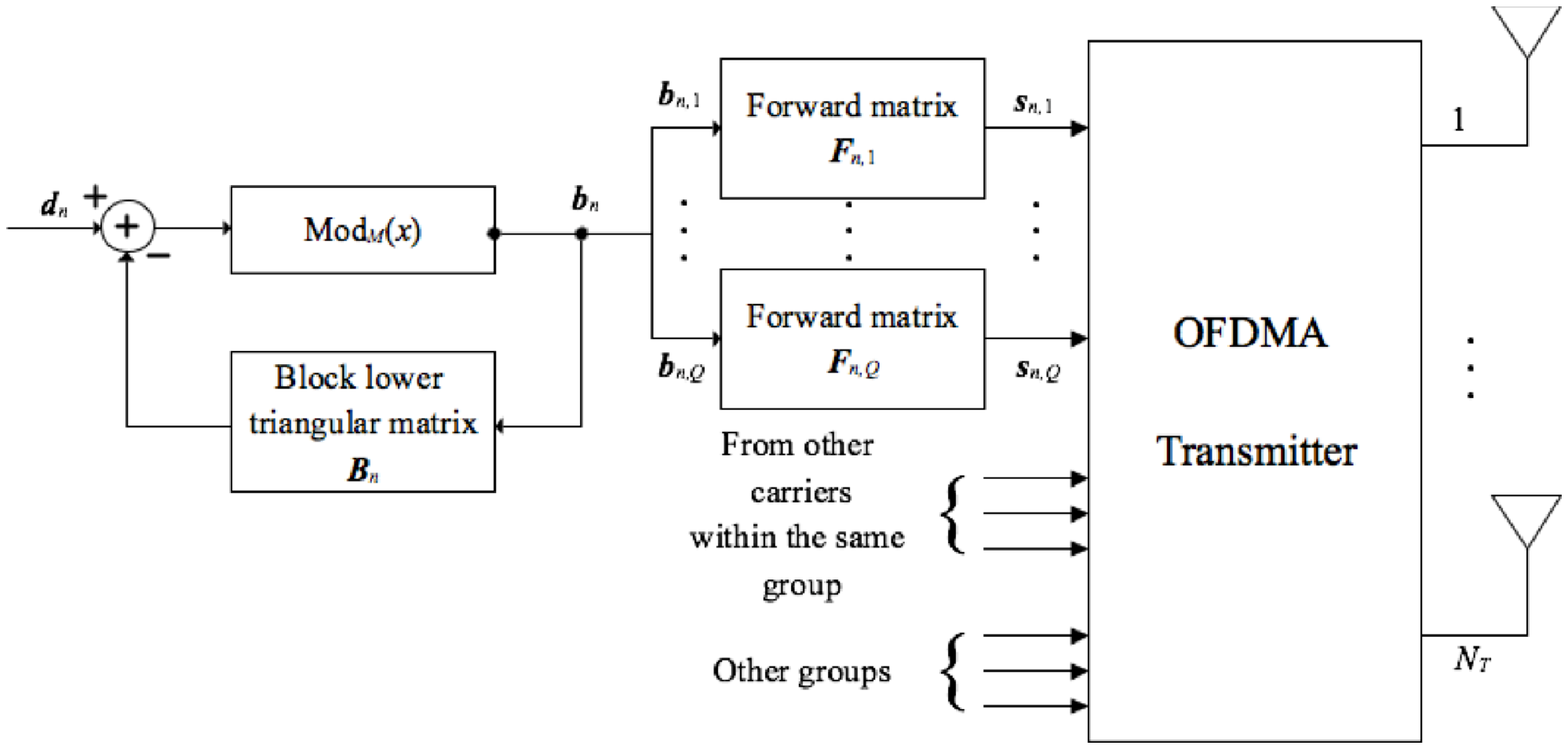}
\end{center}
\caption{Block diagram of the THP technique operating at user level.}
\label{fig:transmitter}
\end{figure}

\begin{figure}[t]
\begin{center}
\includegraphics[width=.75\textwidth]{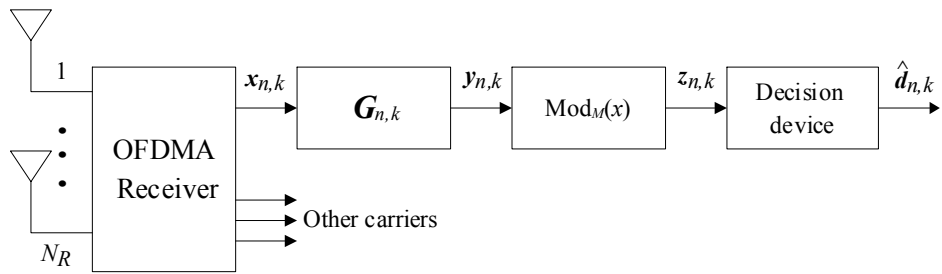}
\end{center}
\caption{Block diagram of the receiver at each MT.}
\label{fig:receiver}
\end{figure}

The $L\le \lfloor N_T/Q\rfloor$ symbols transmitted to the $k$th user over the $n$th subcarrier are denoted by $\left\{ {d_{n,k} (\ell);\;\ell = 1,2, \ldots ,L} \right\}$. They belong to an $M$-ary quadrature-amplitude modulation (QAM) alphabet with variance $\sigma_d^2 = 2(M - 1)/3$ and are stacked in the $L$-dimensional vector ${\bf{d}}_{n,k}$. As depicted in Fig. \ref{fig:transmitter}, the $QL$-dimensional data vector ${\bf{d}}_n = [{\bf{d}}^T_{n,1}, {\bf{d}}^T_{n,2},\ldots,{\bf{d}}^T_{n,Q} ]^T$ is pre-coded in a recursive fashion using a
\emph{strictly block} lower triangular matrix $\mathbf{B}_n \in \CC^{QL \times QL}$ and a
non-linear operator ${\rm{MOD}}_M ( \cdot )$ that constrains the entries of ${\bf{b}}_{n,i} \in \CC^{L\times 1}$ into the
square region $\aleph = \{ { {x^{(R)} + jx^{(I)} } |x^{(R)} ,x^{(I)}
\in ( { - \sqrt M ,\sqrt M }]} \}$. Denoting by $\left[{\mathbf{B}}_n\right]_{i,\ell}\in \mathbb{C}^{L\times L}$ the $(i,\ell)$th block of ${\mathbf{B}}_n$, we have that ${\bf{b}}_{n,i}\in \CC^{L \times 1}$ can be iteratively computed as \cite{FisherBook}
\begin{equation}\label{eq:b_n,k}
{\bf{b}}_{n,i} = {\bf{d}}_{n,i} - \sum\limits_{\ell = 1}^{i - 1} {\left[{\mathbf{B}}_n\right]_{i,\ell} {\bf{b}}_{n,\ell} }+ {\boldsymbol{\varsigma}}_{n,i} \quad
i=1,2,\ldots,{Q}
\end{equation}
where $\left[{\mathbf{B}}_n\right]_{i,\ell}\in \mathbb{C}^{L\times L}$ is the $(i,\ell)$th block of ${\mathbf{B}}_n$, ${\boldsymbol{\varsigma}}_{n,i}$ is defined as ${\boldsymbol{\varsigma}}_{n,i} = 2\sqrt M {\boldsymbol{\xi}}_{n,i}$ and ${\boldsymbol{\xi}}_{n,i} =
\left[ {\xi_{n,i} (1),\xi_{n,i} (2), \ldots , \xi_{n,i} (L)} \right]^T$ with $\xi_{n,i} (\ell)$
complex-valued quantity, whose real and imaginary parts are suitable
integers that reduce $b_{n,i}(\ell)$ to $\aleph$.
%
%
The above equation
indicates that the modulo operator is equivalent
to adding a vector ${\boldsymbol{\varsigma}}_n$ 
to the input data ${\bf{ d}}_n$. This
produces the \emph{modified} data vector
${\bf{v}}_n={\bf{ d}}_n + {\boldsymbol{\varsigma}}_n = [ {{\bf{v}}_{n,1}^T, {\bf{v}}_{n,2}^T,
,\ldots, {\bf{v}}_{n,Q}^T }]^T$ from which ${\bf{ b}}_n $ is obtained as follows ${\bf{ b}}_n = {\bf{C}}_n^{ - 1} {\bf{v}}_n$ where $\mathbf{C}_n \in \CC^{LQ \times QL}$ is a block \emph{unit-diagonal} and lower triangular matrix given by ${\bf{C}}_n = {\bf{B}}_n + {\bf{I}}_{LQ}$. The pre-coded vectors ${\bf{b}}_{n,i} \in \CC^{L \times 1}$ are then
linearly processed through the \emph{forward} transmit matrices $\mathbf{F}_{n,i} \in \CC^{N_T \times L}$ to produce ${\bf{s}}_{n,i} =\mathbf{F}_{n,i}\mathbf{b}_{n,i}$.
The vectors ${\bf{s}}_{n,i}$ for $n=1,2,\ldots,N$ and $i=1,2,\ldots, Q$ are finally fed to the OFDMA modulator 
and transmitted over the channel using the $N_T$ antennas of the BS array. {As depicted in Fig. \ref{fig:receiver}, at the MT the incoming waveforms are implicitly combined by the receive antennas and passed to an OFDMA demodulator whose outputs take the form in \eqref{eq:x_n,k.0} with $\mathbf{s}_{n,i} = \mathbf{F}_{n,i}\mathbf{b}_{n,i}$.}
The complete elimination of ${\bf{H}}_{n,k} \sum\nolimits_{i=k+1}^{Q}\mathbf{F}_{n,i}\mathbf{b}_{n,i}$ at the transmitter can be achieved by constraining ${\bf{F}}_{n,k}$ to lie in the null space of ${\bf{\bar H}}_{n,k} = [{\bf{H}}_{n,1}^T,{\bf{H}}_{n,2}^T,\ldots,{\bf{H}}_{n,k-1}^T]^T$.
Accordingly, this amounts to letting ${\bf{F}}_{n,k}$ have the following structure 
\begin{equation}\label{eq:F_n,k}
{\bf{F}}_{n,k} = {\bf{V}}^{(0)}_{\bar H_{n,k}}{\bf{U}}_{n,k}
\end{equation}
where ${\bf{U}}_{n,k} \in \CC^{ [N_T - (k-1)N_R ]\times L}$ is an arbitrary matrix and ${\bf{V}}^{(0)}_{\bar H_{n,k}} \in \CC^{N_T \times [N_T - (k-1)N_R]}$ is a matrix whose columns form a basis for the \emph{null space} of ${\bf{\bar H}}_{n,k}$ obtained from its singular value decomposition (SVD).
Setting ${\bf{F}}_{n,k}$ as in \eqref{eq:F_n,k} into (\ref{eq:x_n,k.0}) 
and stacking the received signals of all users into a single vector $
{\bf{x}}_{n} = [ {{\bf{x}}_{n,1}^T {\bf{x}}_{n,2}^T \cdots {\bf{x}}_{n,Q}^T}]^T$, we may write
\begin{equation}\label{eq:x_n.1}
\mathbf{x}_{n} = \mathbf{T}_{n}\mathbf{b}_{n}
+\mathbf{w}_{n}
\end{equation}
where $
{\bf{w}}_{n} = [ {{\bf{w}}_{n,1}^T, {\bf{w}}_{n,2}^T, \ldots, {\bf{w}}_{n,Q}^T}]^T$ and ${\bf{T}}_n \in \CC^{ N_R Q\times QL}$ is a block lower triangular matrix with blocks ${\left[{\bf{T}}_n\right]_{k,i} } \in \CC^{ N_R \times L} $ given by ${\left[ {\bf{T}}_n \right]_{k,i}} = {\bf{H}}_{n,k}{\bf{V}}^{(0)}_{\bar H_{n,i}}{\bf{U}}_{n,i}$
for $k \ge i$.
We are now left with the problem of removing the interference term ${\bf{H}}_{n,k}\sum\nolimits_{i =1}^{k-1} {\bf{V}}^{(0)}_{\bar H_{n,i}}{\bf{U}}_{n,i} {\bf{b}}_{n,i}$ in \eqref{eq:x_n,k.0}. To this end, we decompose ${\bf{T}}_n$ in \eqref{eq:x_n.1} as $\mathbf{T}_{n} = \mathbf{D}_{n}\mathbf{L}_{n}$ 
where $\mathbf{D}_{n} = {\rm{blkdiag}}\{{\left[{\bf{T}}_n\right]_{1,1} },{\left[{\bf{T}}_n\right]_{2,2} }, \ldots,$ ${\left[{\bf{T}}_n\right]_{Q,Q} }\} $
and $\mathbf{L}_{n}$ is a block unit-diagonal and lower triangular matrix 
with 
\begin{equation}\label{eq:E_n.1}
{\left[{\bf{L}}_n\right]_{k,i} }= {\left[{\bf{T}}_n\right]^H_{k,k} } \left({\left[{\bf{T}}_n\right]_{k,k} }{\left[{\bf{T}}_n\right]^H_{k,k} }\right)^{-1}{\left[{\bf{T}}_n\right]_{k,i} }\end{equation}
for $k > i$. Substituting $\mathbf{T}_{n} = \mathbf{D}_{n}\mathbf{L}_{n}$ into \eqref{eq:x_n.1} and recalling that ${\bf{ b}}_n = {\bf{C}}_n^{ - 1} {\bf{v}}_n$ yields $\mathbf{x}_{n} = \mathbf{D}_{n}\mathbf{L}_{n}{\bf{C}}_n^{ - 1} {\bf{v}}_n
+\mathbf{w}_{n}$ 
from which setting ${\bf{C}}_n = \mathbf{L}_{n}$ we obtain $\mathbf{x}_{n} = \mathbf{D}_{n} {\bf{v}}_n+\mathbf{w}_{n}$.
Recalling that $\mathbf{D}_{n}$ has a block-diagonal structure with blocks given by ${\left[ {\bf{T}}_n \right]_{k,k}} = {\bf{H}}_{n,k}{\bf{V}}^{(0)}_{\bar H_{n,k}}{\bf{U}}_{n,k}$, it follows that the multi-user MIMO system has been decoupled into $\left| \mathcal K_n \right|$ parallel
single-user MIMO links given by 
\begin{equation}\label{4.0}
\mathbf{x}_{n,k} = \mathbf{H}^\prime_{n,k} {\bf{U}}_{n,k}{\bf{v}}_{n,k}
+\mathbf{w}_{n,k}
\end{equation}
each of which represented by the \emph{equivalent} channel transfer matrix $\mathbf{H}^\prime_{n,k}= {\bf{H}}_{n,k}{\bf{V}}^{(0)}_{\bar H_{n,k}}$. This means that each user may operate in its corresponding link independently without affecting the other active users. Henceforth, we denote by $\mathbf{H}^\prime_{n,k} = {\boldsymbol{\Omega}}_{H^{\prime}_{n,k}}{\boldsymbol{\Lambda}}^{1/2}_{H^{\prime}_{n,k}} {\bf{V}}^{(1)^H}_{H^{\prime}_{n,k}}$
the SVD of $\mathbf{H}^\prime_{n,k}$. As mentioned before, the vectors $\{\mathbf{x}_{n,k}\}$ are processed by the $k$th mobile terminal for
data recovery.

\vspace{-0.4cm}
\subsection{User partitioning}
\vspace{-0.1cm}
{As mentioned above, MAI mitigation in SDMA-OFDMA systems is accomplished not only by precoding the users' data but also by partitioning the users  and dynamically  assigning the radio channels. Unfortunately, optimal grouping is a problem of combinatorial complexity whose solution can only be found through an exhaustive search. To overcome this problem, a heuristic approach widely used in the literature is to partition users on the basis of their space cross-correlations (see for example \cite{MaKlei2010}). Although reasonable, this approach has still a large complexity as it requires the calculation of the cross-correlations among all users in the system over all available channels.  Alternatively, in this work we exploit the fact that THP can be viewed as the transmit counterpart of the vertical Bell Labs layered space-time (V-BLAST) architecture and thus we order the users according to their channel qualities  
as as originally proposed in \cite{Wolniansky1998} and later extended to THP in \cite{Kusume2005}. In our context, the channel quality of the $k$th user is measured by the following quantity:}
\begin{equation}
\pi(k)= \frac{1}{N}\sum\limits_{n=1}^{N}\tr\left(\mathbf{H}_{n,k}^{H}\mathbf{H}_{n,k}\right) = {\frac{1}{N}\sum\limits_{n=1}^{N}\sum\limits_{\ell=1}^{L}\lambda_{{H}_{n,k}}(\ell)}
\label{eq: metric}
\end{equation}
{where $\{\lambda_{{H}_{n,k}}(\ell)\}$ denote the eigenvalues of $\mathbf{H}_{n,k}^{H}\mathbf{H}_{n,k}$.
The above quantities are used to partition users according to a \emph{worst-first} criterion. In doing so, the users with the most attenuated channels are allocated in set $\mathcal S^{(1)}$ whereas the users with the best channels are grouped in $\mathcal S^{(Q)}$.
This choice is motivated by the fact that the null-space projection in \eqref{eq:F_n,k} progressively reduces the available spatial diversity as the group index tends to $Q$ and the number of rows of ${\bf{\bar H}}_{n,k}$ increases up to $\left(Q-1\right)N_{R}$. Therefore, since power consumption is in general dominated by users with the worst channel conditions, we give those users higher priority by placing them in set $\mathcal S^{(1)}$. Observe that the MAI arising among users (in different sets) allocated on the same subcarriers is mitigated jointly by THP and dynamic channel assignment. With the objective of minimizing the overall required power, channel assignment will automatically couple users that tend to not interfere with each other.
It is worth observing that the same ordering strategy is used in \cite{Moretti2013} following a different line of reasoning.}
\vspace{-0.3cm}
\section{Linear programming subcarrier assignment}\label{LP}
\vspace{-0.2cm}
Without loss of generality, we focus on the resource allocation problem over the $K/Q$ users within the set $\mathcal S^{(q)}$. For notational convenience, we denote by $\mathbf{a}^{(q)}$ and $\mathbf{U}^{(q)}$ the vector and the matrix obtained stacking the allocation variables and the precoding matrices of the users in $\mathcal S^{(q)}$, respectively. As before, the user {indexes} $\mu_n(i)$ will be relabelled according to the map $\mu_n(i) \leftarrow i$. To make the problem mathematically tractable, we assume also that the precoded symbols $\mathbf{b}_{n,k}$ are statistically independent and with the same power of user data\footnote{Although not rigorously true, this assumption is reasonable for large $M-$QAM constellations with size $M\ge 16$ \cite{FisherBook}.}, i.e., ${\rm{E}}\{\mathbf{b}_{n,k}\mathbf{b}_{n,k}^H\} =\sigma_d^2\mathbf{I}_L$. In these circumstances, using \eqref{eq:F_n,k} it follows that the power required by the BS to transmit the signal $\mathbf{s}_{n,k}$ is given by ${\rm{E}}\{\mathbf{s}_{n,k}\mathbf{s}_{n,k}^H\} = \sigma_d^2{\rm{tr}}\{\mathbf{U}_{n,k}^H\mathbf{U}_{n,k}\}$.
%
%
The optimization problem can be thus mathematically formulated as:
\begin{align}
\mathop {\min }\limits_{\mathbf{U}^{(q)},\mathbf{a}^{(q)}} & \sum\limits_{n=1}^N \quad\sum\limits_{k \in \mathcal S^{(q)}} a_{n,k}{\rm{tr}}\left\{\mathbf{U}_{n,k}^H\mathbf{U}_{n,k}\right\} \label{opt_problem}\\
\st \quad & \sum\limits_{n=1}^{N} a_{n,k}\sum\limits_{\ell=1}^L {\rm{MSE}}_{n,k} (\ell) \le \gamma_k \quad k \in \mathcal S^{(q)} \tag{\ref{opt_problem}.1} \label{opt_problem.1} \quad \text{and} \quad \sum\limits_{n=1}^Na_{n,k}\ge n_k \quad k \in \mathcal S^{(q)} \nonumber 
\end{align}
which is a mixed-integer non-linear problem and thus not convex and very difficult to solve. A possible way out is to decouple the power allocation and subcarrier assignment problems. This can be achieved by assigning $n_k$ subcarriers to the $k$th user and designing the processing matrices such that the following constraint is satisfied
\begin{equation}\label{6.1}
\sum\limits_{\ell=1}^L {\rm{MSE}}_{n,k} (\ell) \le \frac{\gamma_k}{n_k}.
\end{equation}
In this framework, the power is no longer an optimization variable but simply the cost of using $n_k$ subcarriers \cite{Moretti2011}. In particular, the cost $c_{n,k}$ of using subcarrier $n$ for user $ k \in \mathcal S^{(q)}$ {can be computed as}
\begin{align}
\mathop {\min }\limits_{\;\;\mathbf{U}_{n,k}} \quad {\rm{tr}}\left\{\mathbf{U}_{n,k}^H\mathbf{U}_{n,k}\right\} \quad
\st \quad \sum\limits_{\ell=1}^L {\rm{MSE}}_{n,k} (\ell) \le\frac{ \gamma_k}{n_k}. 
\label{powProb}
\end{align}
%
%
Once the solution of \eqref{powProb} is obtained, \eqref{opt_problem} can be recast as a linear integer programming (LIP) problem:
\begin{align}\label{6.13}
 \mathop {\min }\limits_{{\mathbf{a}}^{(q)}} & \quad\sum\limits_{n=1}^N\sum\limits_{k \in \mathcal S^{(q)}} a_{n,k}c_{n,k}\\ \nonumber \st & \quad \sum\limits_{n = 1}^N a_{n,k} = n_k \quad k \in \mathcal S^{(q)} \quad \text{and} \quad \sum\limits_{k \in \mathcal S^{(q)}}a_{n,k} \le 1 \quad \forall n \nonumber 
\end{align}
where the objective function and the constraints are linear in $\{a_{n,k} \}$.
In general, the solution of LIP problems can be found either performing an exhaustive search or relaxing the integrality condition on the allocation variable. In this particular case, the channel assignment in \eqref{6.13} has the advantage that can be modelled as a \emph{minimum cost flow} problem and as such it is possible to show that the solution obtained by relaxing the integral condition is the optimal \emph{integral} solution, so that very efficient solvers can be employed with no performance degradation \cite{Moretti2011}. 

\vspace{-0.4cm}
\subsection{Receiver design}\label{Lin}
\vspace{-0.1cm}

{To keep the complexity of the MTs at a tolerable level, we assume that a linear receiver is used for data recovery. As depicted in Fig. \ref{fig:receiver}, vector ${\bf{x}}_{n,k}$ in \eqref{4.0} is first processed by ${\bf{G}}_{n,k} \in \CC^{L\times N_R}$ to obtain 
\begin{align}\label{y_{n,k}}
\mathbf{y}_{n,k} = \mathbf{G}_{n,k}\mathbf{H}^\prime_{n,k} {\bf{U}}_{n,k}{\bf{v}}_{n,k}+\mathbf{G}_{n,k}\mathbf{w}_{n,k}
\end{align}
and then passed to the same modulo operator employed at the transmitter so as to remove the effect of ${\boldsymbol{\varsigma}}_{n,k}$. The output $\mathbf{z}_{n,k} = [z_{n,k}(1),z_{n,k}(2),\ldots,z_{n,k}(L)]^T$ is finally fed to a threshold unit which
delivers an estimate of $\mathbf{d}_{n,k}$. From \eqref{y_{n,k}}, it follows that the received samples depend on $\mathbf{G}_{n,k}$ and ${\bf{U}}_{n,k}$. The latter must be designed so as to mitigate co-channel interference while satisfying the QoS constraints. For this purpose, we adopt a ZF approach in which multi-stream interference is completely eliminated and the remaining degrees of freedom are exploited to minimize the power consumption under the constraint on the MSEs. The complete elimination of the multi-stream interference implies that 
\begin{equation}\label{C.2}
{\bf{G}}_{n,k} {\bf{H}}^{\prime}_{n,k} {\bf{U}}_{n,k} = {\bf{I}}_{L}.
\end{equation}
In these circumstances, the output $z_{n,k}(\ell)$ from the
modulo operator takes the form\footnote{{In writing $z_{n,k}(\ell)  = d_{n,k}(\ell)  + {{n_{n,k}(\ell) }}$, we have neglected for simplicity the modulo-folding effect on the thermal noise. Although not rigorous, this assumption is quite reasonable for moderate values of signal-to-noise ratios (see for example the book of Robert F. H. Fisher \cite{FisherBook} for a complete treatment of the subject).}} $z_{n,k}(\ell)  = d_{n,k}(\ell)  + {{n_{n,k}(\ell) }}$ and its corresponding MSE results given by ${\rm{MSE}}_{n,k} (\ell) = {\sigma^2 [{\bf{G}}_{n,k}{\bf{G}}^H_{n,k}]_{\ell,\ell}}.
$}
It can be shown that the optimal ${\bf{G}}_{n,k}$ satisfying \eqref{C.2} and minimizing each ${\rm{MSE}}_{n,k} (\ell)$ is the minimum norm solution of \eqref{C.2} \cite{KayBook}. The latter is found to be ${\bf{G}}_{n,k} = ({\bf{ U}}^H_{n,k}{\bf{H}}^{\prime^H}_{n,k}{\bf{ H}}^{\prime}_{n,k}{\bf{U}}_{n,k} )^{ - 1} {\bf{ U}}^H_{n,k}{\bf{H}}^{\prime^H}_{n,k}$
from which it follows that ${\rm{MSE}}_{n,k} (\ell) = {\sigma^2 [({\bf{ U}}^H_{n,k}{\bf{H}}^{\prime^H}_{n,k}{\bf{ H}}^{\prime}_{n,k}{\bf{U}}_{n,k} )^{ - 1}]_{\ell,\ell}}$.
We now proceed with the design of the matrix ${\bf{U}}_{n,k}$, which requires to solve the following problem:
\begin{align}\label{C.6}
 \mathop {\min }\limits_{\{\mathbf{U}_{n,k}\}}  \quad {\rm{tr}}\left\{\mathbf{U}_{n,k}^H\mathbf{U}_{n,k}\right\} \quad \st \quad \sum\limits_{\ell=1}^L {\sigma^2 \left[\left({\bf{ U}}^H_{n,k}{\bf{H}}^{\prime^H}_{n,k}{\bf{ H}}^{\prime}_{n,k}{\bf{U}}_{n,k} \right)^{ - 1}\right]_{\ell,\ell}}\le \frac{\gamma_k}{n_k}.
\end{align}
The solution can be computed as follows.
\begin{proposition}
The optimal ${\bf{U}}_{n,k} $ in \eqref{C.6} takes the form 
\begin{equation}\label{C.9}
{\bf{U}}_{n,k} ={\bf{V}}^{(1)}_{H^{\prime}_{n,k}}{\mathbf{\Lambda}}_{U_{n,k}}^{{1/2}}\mathbf{S}_{n,k}^H 
\end{equation}
where ${\bf{V}}^{(1)}_{H^{\prime}_{n,k}}$ is obtained from the SVD of $\mathbf{H}^\prime_{n,k}$, $\mathbf{\Lambda}_{U_{n,k}}$ is diagonal and $\mathbf{S}_{n,k} \in \CC^{L \times L}$ is a suitable unitary
matrix such that ${\rm{MSE}}_{n,k} (\ell) = \epsilon_k$ for $\ell = 1, 2, \ldots,L$
with $\epsilon_k = \frac{1}{L}\frac{\gamma_k}{n_k}$.
In addition, the diagonal elements of $\mathbf{\Lambda}_{U_{n,k}}$ are given by
\begin{equation}\label{C.12}
\lambda_{U_{n,k}}(\ell) = \sqrt{\nu_{n,k}\frac{\sigma^2}{\lambda_{H^{\prime}_{n,k}}(\ell)}} \quad \ell=1,2,\ldots,L
\end{equation}
where ${\nu_{n,k}}$ is such that $\sum\nolimits_{\ell =1}^L\frac{\sigma^2}{\lambda_{U_{n,k}}(\ell) \lambda_{H^{\prime}_{n,k}}(\ell)}=\frac{\gamma_k}{n_k}$.
\end{proposition}

\emph{Proof}: The proof is omitted for space limitations but it can be derived using the results illustrated in \cite{SanguinettiTVT2012} since the sum of the MSEs is a Schur-convex function.\hfill$\blacksquare$

Using the results of Proposition 1, the cost $c_{n,k}$ in \eqref{6.13} is eventually given by 
\begin{equation}
\label{costiLin}
c_{n,k} = \sum\limits_{\ell=1}^L \lambda_{U_{n,k}}(\ell)
\end{equation}
with $\lambda_{U_{n,k}}(\ell)$ computed as in \eqref{C.12}.

\vspace{-0.5cm}
\subsection{Complexity analysis}\label{NL}\vspace{-0.2cm}
{All the operations required by the proposed solution are summarized in Algorithm \ref{AlgI} whose computational load can be assessed in terms of the number of required floating point operations (flops) as follows\footnote{{In doing so, we make use of the following results: \emph{i}) the multiplication of $\mathbf{A}\in \mathbb{C}^{m\times n}$ and $\mathbf{B}\in \mathbb{C}^{n\times p}$ requires $\mathcal{O}(2mnp)$ flops; \emph{ii}) evaluating the SVD of $\mathbf{A}\in \mathbb{C}^{m\times n}$ needs $\mathcal{O}(mn^2)$ flops; \emph{iii}) the inverse of $\mathbf{A}\in \mathbb{C}^{n\times n}$ requires $\mathcal{O}(n^3)$ flops.}}. Observe that computing the quantities $\{\pi(k)\}$ requires $\mathcal{O}(NKN_TN_R)$ flops whereas computing the power cost $c_{n,k}$ according to \eqref{costiLin} basically requires first to evaluate the SVDs of ${\bf{\bar H}}_{n,k}$ for $k=1,2,\ldots, K/Q$ and $n=1,2,\ldots,N$ and then those of ${\bf{H}}^\prime_{n,k}$ in \eqref{4.0} for $k=1,2,\ldots, K$ and $n=1,2,\ldots,N$. The total number of flops required for these two operations are summarized in the second and third row of Table I. In writing these figures, we have taken into account that evaluating the SVDs of ${\bf{H}}^\prime_{n,k}$ requires {$\mathcal{O}(Q/2(Q-1)N_TN_R^2 + QN_RN_T^2 )$} flops in total since $\mathcal{O}({Q}N_RN_T^2)$ flops are needed to compute $\mathbf{H}^\prime_{n,k}= {\bf{H}}_{n,k}{\bf{V}}^{(0)}_{\bar H_{n,k}}$ whereas $\mathcal{O}(Q/2(Q-1)N_TN_R^2)$ flops are required for the SVD. Summing all the above terms it turns out that the overall complexity for computing all costs $\{c_{n,k}\}$ is approximately given by $\mathcal{O}(NKQN_RN_T^2)$. The complexity of solving \eqref{6.13} is an open research issue. The latest results (see for example \cite{HuaJeb11} references therein) place the complexity of the assignment problem in a range between $\mathcal{O}\left(\kappa^{2}\right)$ and $\mathcal{O}\left(\kappa^{2.5}\right)$ with $\kappa$ being the total number of nodes. In our case, the number of nodes is the sum of the number of users per single allocation problem plus the number of subcarriers, i.e., $\kappa = N+K/Q$. Since we have $Q$ distinct subproblems to solve, the overall complexity of the LP optimization is approximately given by $\mathcal{O}\left(Q(N+K/Q)^{2.5}\right)$ flops. The computation of ${\bf{B}}_{n} = {\bf{C}}_{n} - {\bf{I}}_{LQ}$ in \eqref{eq:b_n,k} with ${\bf{C}}_{n} = {\bf{L}}_{n}$ can be assessed as follows. Evaluating each ${\left[{\bf{L}}_n\right]_{k,i} }$ in \eqref{eq:E_n.1} requires $\mathcal{O}((N_R^3 + 4LN_R^2))$ flops. Since the total number of matrices ${\left[{\bf{L}}_n\right]_{k,i} }$ is $Q/2(Q-1)$, it follows that $\mathcal{O}(NQ/2(Q-1)(N_R^3 + 4LN_R^2))$ flops are required to obtain all matrices $\{{\bf{C}}_{n}\}$ and thus  all $\{{\bf{B}}_{n}\}$. {The computational load for obtaining $\{\mathbf{F}_{n,k}\}, \{\mathbf{G}_{n,k}\}$ and $\{\mathbf{U}_{n,k}\}$ can be reasonably neglected as it basically require to put together all the unitary matrices computed above with SVDs.} The processing requirements of the proposed two-layer architecture are summarized in Table I from which it follows that the overall number of flops is approximately given by $\mathcal{O}\left(Q(N+K/Q)^{2.5} + NKQN_RN_T^2 + NQ^2N_R^3\right)$. The latter is comparable to the computational load required by the scheme illustrated in \cite{Moretti2013} as it is dominated by the computational burden required by the LP approach, especially when the number of subcarriers relatively large. However, as shown in the sequel, the proposed solution provides much better performance in terms of power reduction with respect to \cite{Moretti2013} thanks to the underlying THP scheme.}


\vspace{-0.3cm}
\section{Numerical results}
\vspace{-0.2cm}
We consider a system with $K$ uniformly distributed users in a cell of radius {$R = 100$} m. The propagation channel is static, frequency-selective and modelled as a Rayleigh fading process with an exponentially decaying power delay profile. The path loss exponent is {$\beta = 4$}. Unless noted differently, the number of users is $K=16$. 
\par 
We compare the proposed architecture, denoted by THP Tx - Lin Rx, with three other algorithms: \emph{a}) a ZF linear beam-former, denoted as ZF Tx, \emph{b}) a THP scheme, denoted as THP Tx (see for example \cite{Zhou2006c}), and \emph{c}) the architecture proposed in \cite{Moretti2013} that employs linear processing at both the transmitter and the receiver (Lin Tx - Lin Rx). In details, letting ${\bf{H}}_{n} = [{\bf{H}}^T_{n,1} \, {\bf{H}}^T_{n,2}\, \cdots {\bf{H}}^T_{n,Q}]^T$ and ${\bf{F}}_{n} = [{\bf{F}}_{n,1} \, {\bf{F}}_{n,2}\, \cdots {\bf{F}}_{n,Q}]^T$, the precoding matrix for ZF Tx is
$\mathbf{F}_n=\mathbf{H}_n^H(\mathbf{H}_n\mathbf{H}_n^H)^{-1}$. The THP Tx architecture is realized by setting $\mathbf{F}_n=\mathbf{Q}_n$ and $\mathbf{C}_n = \mathbf{R}_n^{-H}$ with $\mathbf{Q}_n$ and $\mathbf{R}_n$ being computed as the QR decomposition $\mathbf{H}_n^H$, i.e., $\mathbf{H}_n^H= \mathbf{Q}_n\mathbf{R}_n$. Both ZF Tx and THP Tx schemes are designed to remove the inter-stream and  inter-user interference at the transmitter  so that the receive filter is $\mathbf{G}_{n,k} = \mathbf{I}_{L}$.

{We consider three different scenarios, summarised in Table II, which are designed to observe the behaviour of the proposed algorithms when the total number of available channels per user is fixed and frequency channels are progressively replaced by  streams in the spatial domain. More in details, the first scenario, referred to as $S^{(1)}$, is a $2 \times 1$ MIMO system with a bandwidth $W^{(1)}=10$ MHz and  $N^{(1)} = 64$ orthogonal subchannels. The bandwidth of Scenario $S^{(2)}$ is  $W^{(2)} =5$ MHz, spanning $N^{(2)}=32$ subchannels with a $4\times 2$ MIMO configuration. Scenario $S^{(3)}$ transmits over a bandwidth $W^{(3)} =2.5$ MHz with $N^{(3)}=16$ subchannels and employs a $8\times 4$ configuration.  For each scenario we assume that the number of allocated subcarriers is $n_{k}^{(i)}=N^{(i)} \times Q/K$ and the total number of channels per user is  $n_{k}^{(i)}L^{(i)}=8$ ($i=1,\dots,3$; $k=1,\dots,K$) regardless of the scenario considered}.

\begin{figure}[t]
\begin{center}
\includegraphics[width=.6\textwidth]{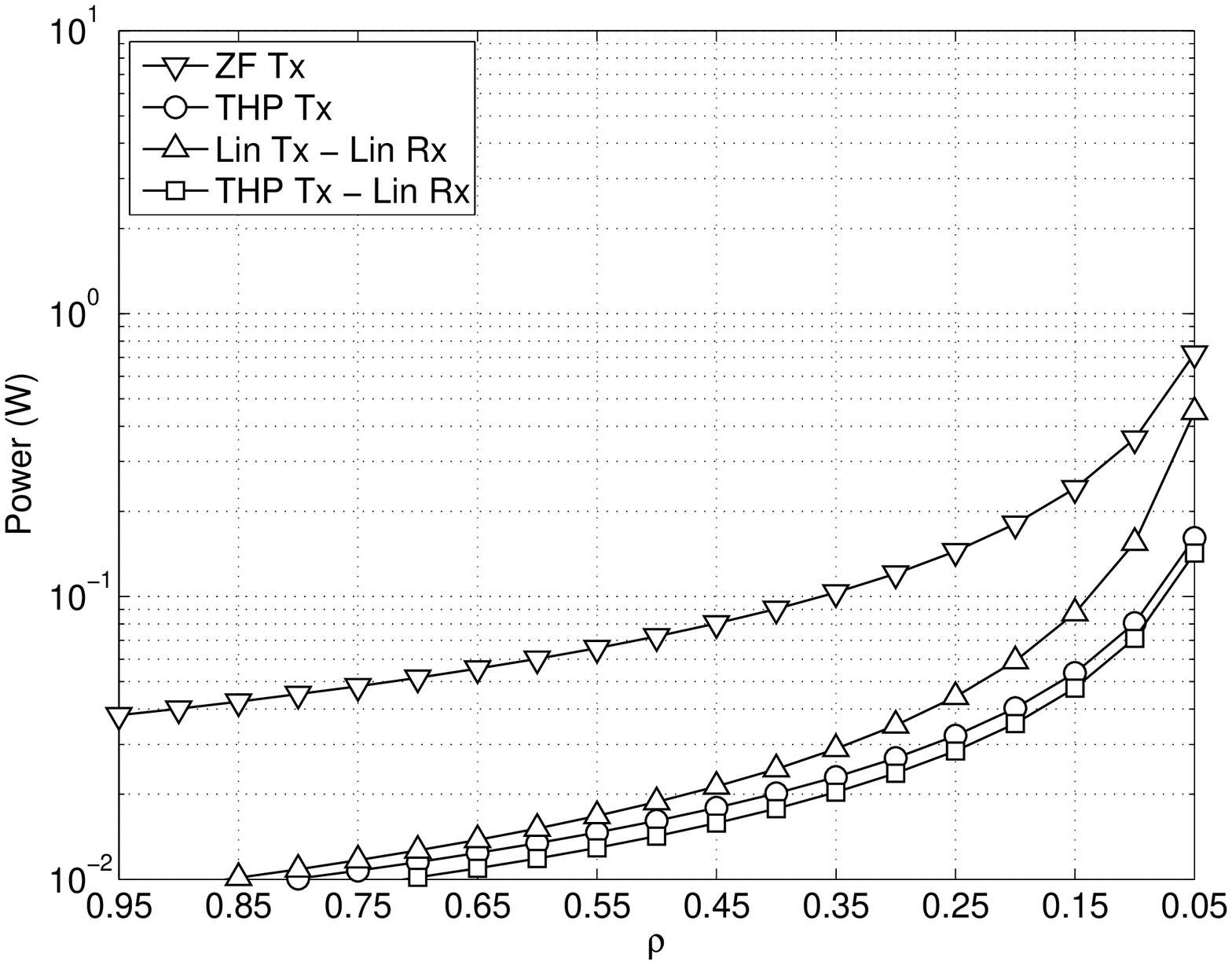}
\end{center}
\caption{Total power consumption for the investigated solutions in scenario $S^{(1)}$ for {different MSEs for data stream}.}
\label{fig:1_1}
\end{figure}

\begin{figure}[t]
\begin{center}
\includegraphics[width=.6\textwidth]{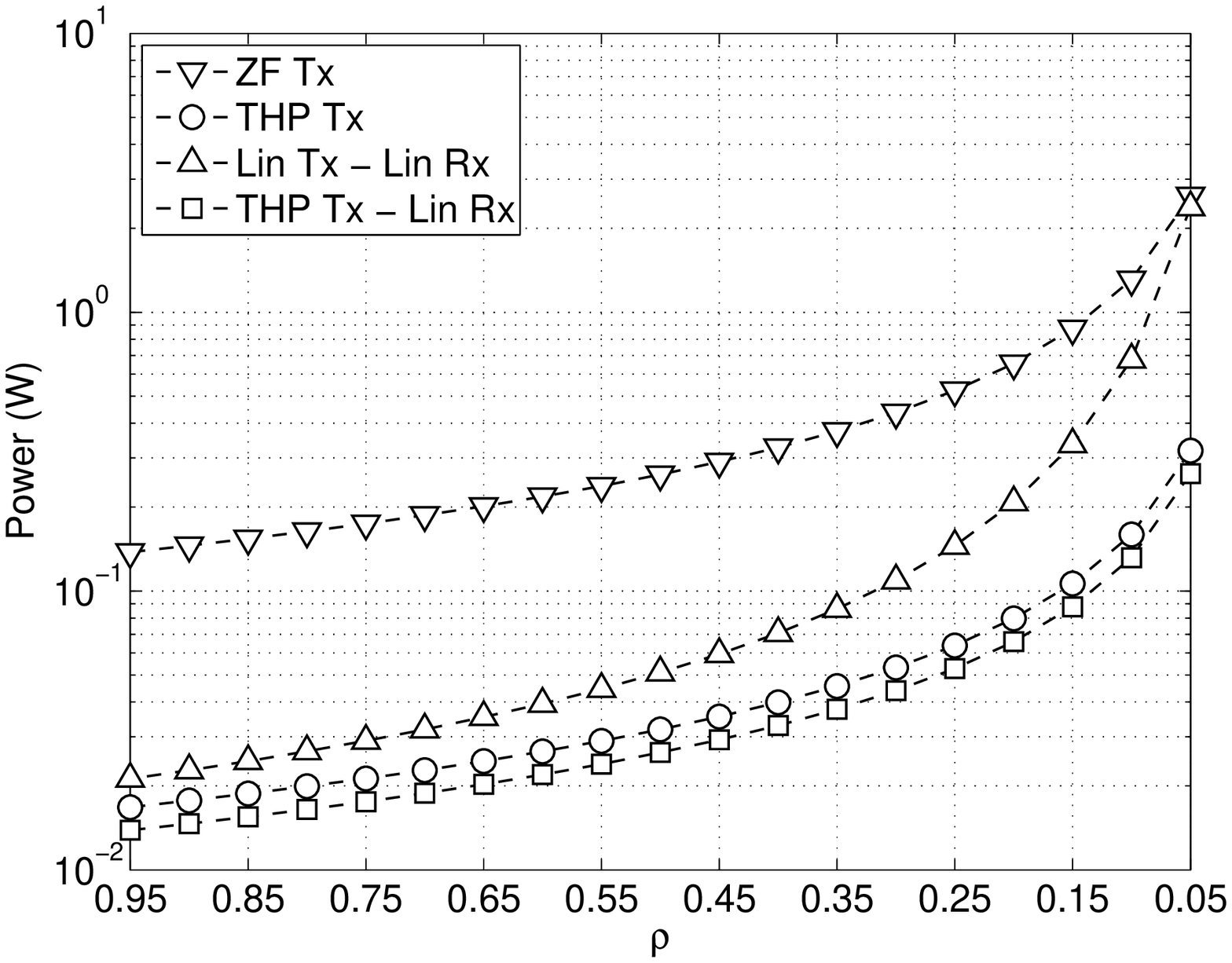}
\end{center}
\caption{Total power consumption for the investigated solutions in scenario $S^{(2)}$ for {different MSEs for data stream}.}
\label{fig:1_2}
\end{figure}

\begin{figure}[t]
\begin{center}
\includegraphics[width=.6\textwidth]{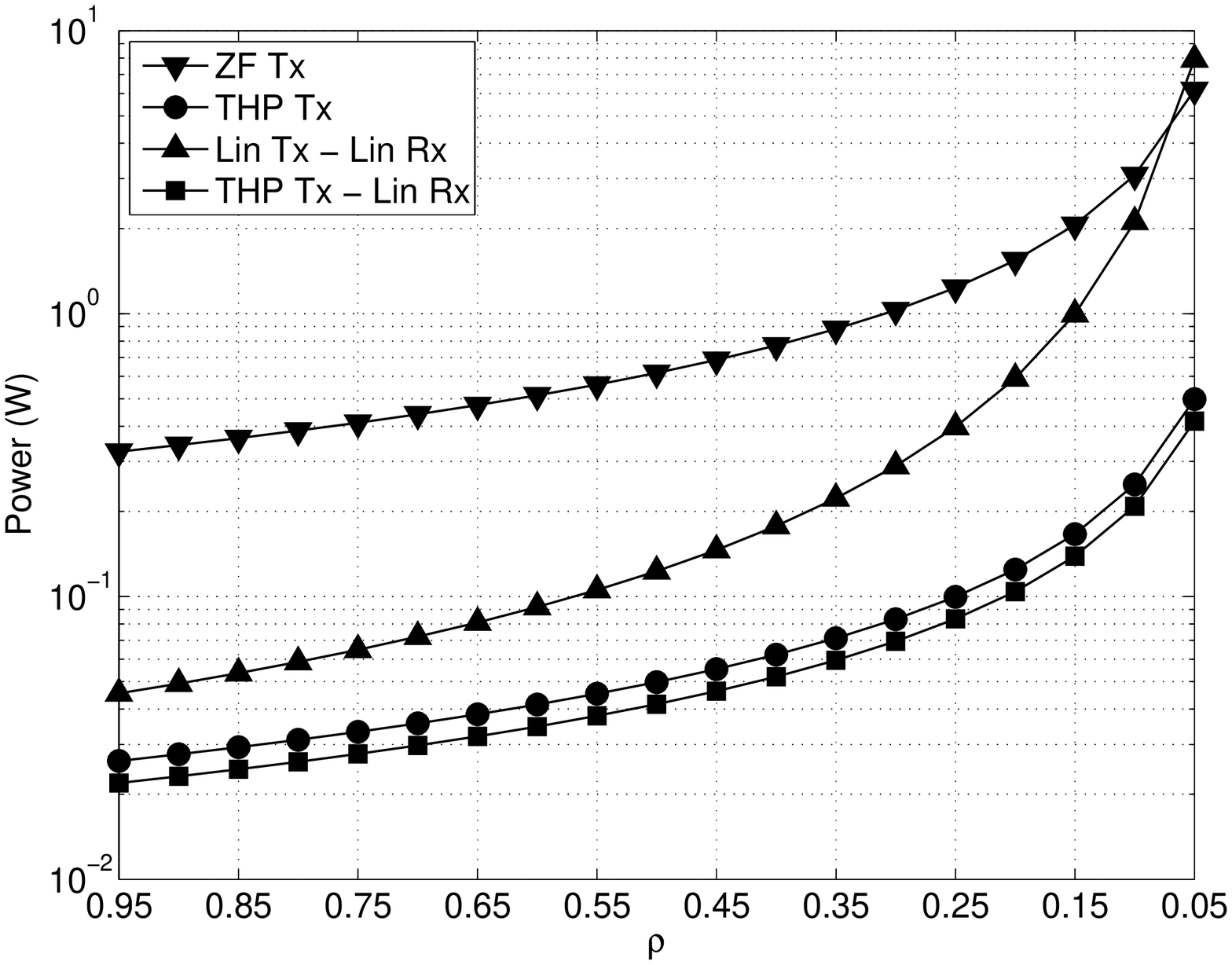}
\end{center}
\caption{Total power consumption for the investigated solutions in scenario $S^{(3)}$ for {different MSEs for data stream}.}
\label{fig:1_3}
\end{figure}

Figs. \ref{fig:1_1} -- \ref{fig:1_3} {report the total transmit power} for the three scenarios as a function of the average target MSE $\rho$ per data stream. By design, for a given value of $\rho$, the overall MSE is $\gamma^{(i)}_{k}=8\rho$ ($i=1,\dots,3$; $k=1,\dots,K$). Results show that the gains obtained thanks to the implementation of non-linear processing  progressively increase from scenario $S^{(1)}$ to $S^{(3)}$, as the spatial dimension becomes more important. 
\par
In particular, Fig. \ref{fig:1_1} shows that, with a $2\times1$ configuration and 64 channels, all the schemes, except ZF Tx, tend to have similar performance. The effect of resource allocation is predominant and the users transmitting on the same channel are sufficiently separated regardless of the specific architecture. 
\par
{As the number of orthogonal frequency channels is reduced, the consequent diminution in frequency diversity is only partially compensated by the larger number of antennas: in facts, even if the total number of channels is the same, the spatial streams tend to be more correlated. In this case, the choice of the transceiver architecture plays a very important role since channel allocation alone is not able to fully exploit all the diversity of the the system. The results plotted in Fig. \ref{fig:1_2} show that the THP-based schemes largely outperform all other solutions.}
\par
{The same trend appears in Fig. \ref{fig:1_3}, where THP Tx - Lin Rx effectively exploits the spatial diversity provided by the multiple antennas. Scenario $S^{(1)}$ requires less power when compared to $S^{(2)}$ and $S^{(3)}$ as it occupies a larger bandwidth. In scenarios $S^{(2)}$ and $S^{(3)}$, the proposed scheme takes advantage of the increased spatial dimension to transmit the same amount of data employing a comparable amount of power and occupying only a fraction of the bandwidth.}

\begin{figure}[t]
\begin{center}
\includegraphics[width=.6\textwidth]{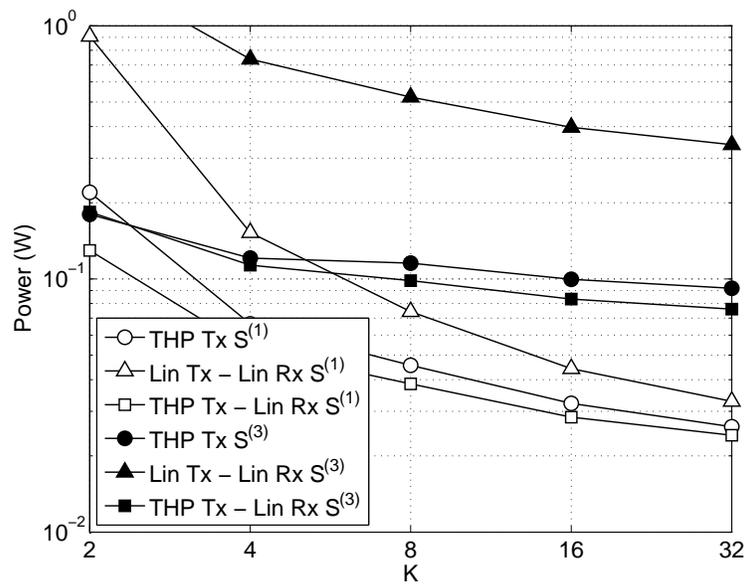}
\end{center}
\caption{Total power consumption for the investigated solutions for different number of users.}
\label{fig2}
\end{figure}

Fig. \ref{fig2} shows {the total transmit power} for an average {target MSE $\rho= 0.25$} as a function of $K$ for $S^{(1)}$ and $S^{(3)}$. For ease of representation, only the results of THP Tx,  Lin Tx - Lin Rx and THP Tx - Lin Rx are reported. As before, the parameters are set so that the number of data stream per user is the same (regardless of the specific scenario). 
{An accurate inspection of the results shows that for scenario $S^{(1)}$, the performance of the three algorithms tend to be very close for $K\geq 16$, when the resource allocation algorithm is able to fully exploit both multi-user and frequency diversity. The situation is remarkably different for scenario $S^{(3)}$ where it appears that resource allocation alone is not sufficient to completely deal with MAI. In fact, all multiuser diversity is already exploited for $K=8$ and further increase of the number of users produce only marginal improvements. In this case, the THP Tx - LIN Rx configuration outperforms the other two schemes thanks to its capability to cancel the MAI.}

\vspace{-0.3cm}
\section{Conclusions}
We have derived a resource allocation scheme for the downlink of SDMA-MIMO-OFDMA systems. {The proposed solution relies on a layered architecture in which MAI is first removed by means of a THP technique operating at user level and then channel assignment and transceiver design are jointly addressed using a ZF-based linear programming approach that aims at minimizing the power consumption while satisfying specific QoS requirements given as the sum of the MSEs over the assigned subcarriers. The proposed approach outperforms the existing solutions,  especially when the frequency diversity is small and the number of spatial modes is large.} 

\vspace{-0.3cm}
\bibliographystyle{IEEEtran}
\bibliography{IEEEabrv,references}

\newpage

{
\begin{algorithm}[t]
	\small \caption{Proposed two-layer architecture} \label{AlgI}
	\begin{algorithmic}[1]
		\For {user $ k=$ 1 to $K$}
    			\State Compute $\pi(k)= \frac{1}{N}\sum\limits_{n=1}^{N}\tr\left(\mathbf{H}_{n,k}^{H}\mathbf{H}_{n,k}\right)$.
		\EndFor
		\State Sort users according to $\pi(k)$ and group them in $Q$ sets $\{\mathcal{S}^{(1)},\dots,\mathcal{S}^{(Q)}\}$.
		\For  {group $i=1$ to $Q$}
    			\For {user $k=1$ to $|\mathcal{S}^{(i)}|$}
				\For {subcarrier $n=1$ to $N$}
					\State Compute the power cost $c_{n,k}$ according to \eqref{costiLin}.	
				\EndFor
			\EndFor
			\State Solve the resource allocation problem in \eqref{6.13}.
					\For {subcarrier $n=1$ to $N$}
				   	\State Compute ${\bf{B}}_{n} = {\bf{C}}_{n} - {\bf{I}}_{LQ}$.
			\EndFor
			\For {user $k=1$ to $|\mathcal{S}^{(i)}|$}
					\For {subcarrier $n=1$ to $N$}
				\State Compute $\{{\bf{F}}_{n,k}$, ${\bf{G}}_{n,k}$, ${\bf{U}}_{n,k}\}$.	
			\EndFor
			\EndFor
			\EndFor
	\end{algorithmic}
\end{algorithm}}

\begin{table}[t] 
\label{TableI}
 \centering
 \caption{Computational load}
 \begin{tabular}{||c|c||}

    \hline
    Operation& Flops\\
     \hline
     \hline
     Computing quantities $\{\pi(k)\}$ & $\mathcal{O}(NKN_TN_R)$\\
     \hline
     Evaluating the SVD of ${\bf{\bar H}}_{n,k}$ & $\mathcal{O}({Q}/2(Q - 1) N_RN_T^2)$\\
    \hline
     Evaluating the SVD of ${\bf{H}}^\prime_{n,k}$ & $\mathcal{O}(Q/2(Q-1)N_TN_R^2 + QN_RN_T^2 )$\\
    \hline
        Solving the LP problem in \eqref{6.13} &$\mathcal{O}\left(Q(N+K/Q)^{2.5}\right)$\\
    \hline
     Computing all matrices $\{\mathbf{B}_n\}$ & $\mathcal{O}(NQ/2(Q-1)(N_R^3 + 4LN_R^2))$\\
    \hline
    \end{tabular}
\medskip
\end{table}

\begin{table}[t] 
\label{TableII}
 \centering
 \caption{Simulation scenarios}
 \begin{tabular}{||c|c|c|c||}

    \hline
     & $S^{(1)}$& $S^{(2)}$&$S^{(3)}$\\
     \hline
     \hline
     MIMO configuration & $2 \times 1$&$4 \times 2$ & $8 \times 4$\\
     \hline
     bandwidth $W^{(i)}$ (MHz) & $10$&$5$&$2.5$\\
    \hline
     $\#$ subcarriers $N^{(i)}$ & $64$&$32$ & $16$\\
    \hline
     $\#$ streams per subcarrier per user $L^{(i)}$ & $1$&$2$&$4$\\
    \hline
    $\#$ subcarriers per user $n_{k}^{(i)}$ & $8$&$4$&$2$\\
    \hline
    \end{tabular}
\medskip
\end{table}

\end{document}